\documentclass[
    ,final            
  ]
  {aipproc}

\layoutstyle{6x9}

\begin{document}

\title{Heavy meson production in hot dense matter}


\author{Laura Tolos$^{1,2}$, Daniel Gamermann$^3$, Carmen Garcia-Recio$^4$, Raquel Molina$^5$, Juan Nieves$^5$, Eulogio Oset$^5$,  Angels Ramos$^3$}{
  address={$^1$Theory Group. KVI. University of Groningen,
Zernikelaan 25, 9747 AA Groningen, The Netherlands \\
$^2$ Instituto de Ciencias del Espacio (IEEC/CSIC), Campus Universitat 
Aut\'onoma de Barcelona, Facultat de Ci\`encies, Torre C5, E-08193 Bellaterra 
(Barcelona), Spain\\
$^3$Departament d'Estructura i Constituents de la Mat\`eria,
Universitat de Barcelona,
Diagonal 647, 08028 Barcelona, Spain\\
$^4$Departamento de F{\'\i}sica At\'omica, Molecular y Nuclear, 
Universidad de Granada, E-18071 Granada, Spain \\
$^5$Instituto de F{\'\i}sica Corpuscular (centro mixto CSIC-UV),
Institutos de Investigaci\'on de Paterna, Aptdo. 22085, 46071, Valencia, Spain}
}


\begin{abstract}
The properties of charmed mesons in dense matter are studied using a unitary coupled-channel approach in the nuclear medium which takes into account Pauli-blocking effects and meson self-energies in a self-consistent manner. We obtain the open-charm meson spectral functions in this dense nuclear environment, and discuss their implications on hidden charm and charm scalar resonances and on the formation of $D$-mesic nuclei at FAIR energies.
\end{abstract}

\maketitle

\section{Introduction}
The modifications of the properties of open and hidden charm mesons in a hot nuclear environment are being the focus of recent analysis. Lately charmed baryonic
resonances have received  a lot of attention motivated by the
discovery of quite a few new states by the CLEO, Belle and BABAR
collaborations \cite{facility00}. Moreover, the behaviour of these resonances in the nuclear medium  and the consequences for open and hidden charm meson in dense matter will be part of the  physics program of the PANDA and CBM experiments at the  future FAIR facility
at GSI \cite{gsi00}. Within this new facility, the GSI programme for in-medium modification of hadrons to the
charm sector will be extended  providing a first insight into the charm-nucleus
interaction.  

The in-medium modification of the properties of open-charm mesons may lead to formation of $D$-mesic nuclei \cite{tsushima99}, and will also affect the renormalization of charm and hidden-charm scalar hadron resonances in nuclear matter, providing information not only about their nature but also about the interaction of the open-charm mesons with nuclei. In the present work we obtain the properties of open-charm mesons in dense matter within a self-consistent approach in coupled channels. We then analyze the effect of those self-energies on the properties of dynamically-generated charm and hidden charm scalar resonances and provide some recent results on $D$-nucleus bound states.

\section{\textbf{$D$} and \textbf{$D^*$} mesons in nuclear medium}

The self-energy in symmetric nuclear matter and, hence, the spectral function for open-charm pseudoscalar ($D$) and vector ($D^*$) mesons is obtained following a self-consistent coupled-channel procedure. The transition potential of the Bethe-Salpeter equation for different isospin ($I$), total angular momentum ($J$) and temperature ($T$) [$T_{D(D^*)N}^{I,J}(P_0,\vec{P},T)$] is derived from effective lagrangians, which will be discussed in the following. The $D$ and $D^*$ self-energies are then obtained summing the transition amplitude for the different isospins over the nucleon Fermi distribution at a given temperature, $n(\vec{p},T)$: 
\begin{eqnarray}
\Pi_D(q_0,{\vec q},T) &=& \int \frac{d^3p}{(2\pi)^3}\, n(\vec{p},T)  \, [\, {T_{DN}}^{(I=0,J=1/2)} +
3 \, {T_{DN}}^{(I=1,J=1/2)}\, ] \nonumber \\ 
\Pi_{D^*}(q_0,\vec{q}\,) &=& \int \frac{d^3p}{(2\pi)^3} \, n(\vec{p},T\,) \,
\, \left [~ \frac{1}{3} \, {T}^{(I=0,J=1/2)}_{D^*N}+
{T}^{(I=1,J=1/2)}_{D^*N}+ \right . \nonumber \\
&&  \left . \frac{2}{3} \,
{T}^{(I=0,J=3/2)}_{D^*N}+ 2 \,
{T}^{(I=1,J=3/2)}_{D^*N}\right ] \  ,
\label{eq:selfd}
\end{eqnarray}
\noindent
where $P_0=q_0+E_N(\vec{p},T)$ and $\vec{P}=\vec{q}+\vec{p}$ are
the total energy and momentum of the meson-nucleon pair in the nuclear
matter rest frame, and ($q_0$,$\vec{q}\,$) and ($E_N$,$\vec{p}$\,) stand  for
the energy and momentum of the meson and nucleon, respectively, in this
frame. The self-energy is determined self-consistently since it is obtained from the
in-medium amplitude which contains the meson-baryon loop function, and this quantity itself
is a function of the self-energy. Then, the meson spectral function  reads
\begin{eqnarray}
S_{D(D^*)}(q_0,{\vec q}, T)= -\frac{1}{\pi}\frac{{\rm Im}\, \Pi_{D(D^*)}(q_0,\vec{q},T)}{\mid
q_0^2-\vec{q}\,^2-m_{D(D^*)}^2- \Pi_{D(D^*)}(q_0,\vec{q},T) \mid^2 } \ .
\label{eq:spec}
\end{eqnarray}

\subsection{SU(4) scheme}
\label{su4}

The $D$ meson spectral function is obtained from the Bethe-Salpeter equation in coupled-channels taking, as bare interaction, a type of broken $SU(4)$ $s$-wave Weinberg-Tomozawa (WT) interaction supplemented by an attractive isoscalar-scalar term and using a cutoff regularization scheme. We fix this cutoff by generating dynamically the $I=0$ $\Lambda_c(2595)$ resonance. A new resonance in $I=1$ channel, $\Sigma_c(2880)$, is generated \cite{LUT06,mizutani06}. The in-medium solution at finite temperature 
incorporates Pauli blocking, baryon mean-field bindings and $\pi$ and $D$ meson self-energies \cite{TOL07}.

\begin{figure}
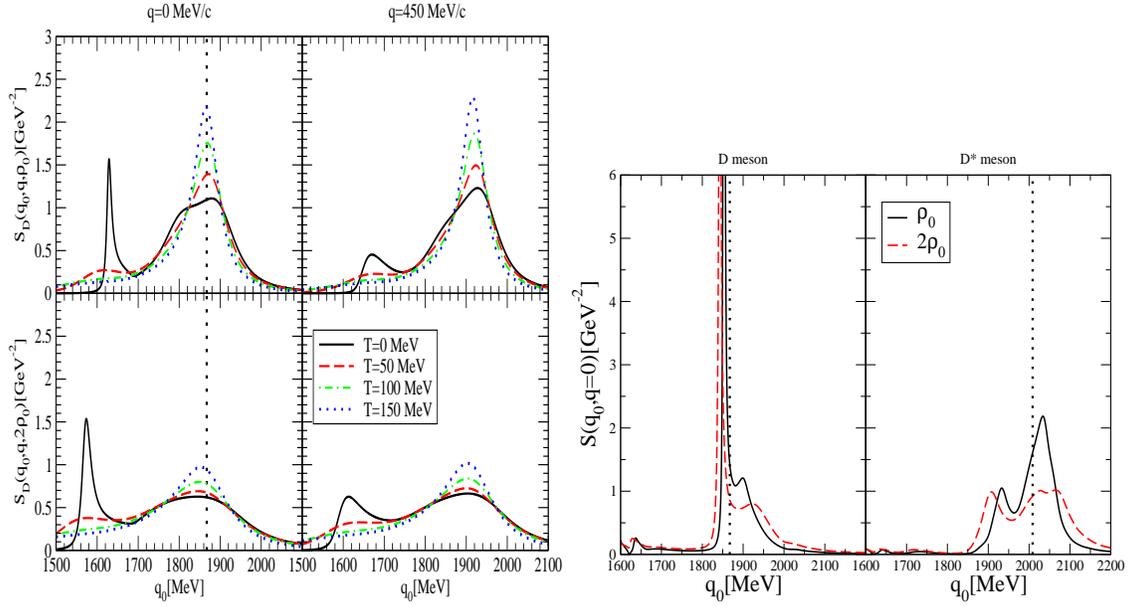

\includegraphics[width=0.5\textwidth, height=8cm]{paper_spectral_tot}
\hfill
\includegraphics[width=0.5\textwidth, height=6cm]{art_spec}
\caption{Left: $D$ meson spectral function for different momenta, temperatures and densities for the SU(4) model. Right: $D$ and $D^*$ spectral functions in nuclear matter at $q=0$ MeV/c and $T$=0 in the SU(8) scheme. We show the $D$ and $D^*$ meson free masses for reference (dotted lines). \label{fig1}}
\end{figure}


In l.h.s. of Fig.~\ref{fig1} we display the $D$ meson spectral function for different momenta, temperatures and densities. At $T=0$ the spectral function shows two peaks. The $\Lambda_c(2595) N^{-1}$ excitation is seen at a lower energy whereas the second one at higher energy corresponds to the quasi(D)-particle  peak  mixed with  the $\Sigma_c(2880) N^{-1}$ state.  Those structures dilute with increasing temperature while the quasiparticle peak gets closer to its free value, becoming narrower as the self-energy
receives contributions from higher momentum $DN$ pairs where the interaction is weaker.
Finite density results in a broadening of the spectral function because of the increased phase space, as previously observed for the $\bar K$ in nuclear matter \cite{Tolos:2008di1,Tolos:2008di2,Tolos:2008di3}.

\subsection{SU(8) model with heavy-quark symmetry}
\label{su8}

Heavy-quark symmetry (HQS) is a QCD spin  symmetry
 that treats on equal footing heavy pseudoscalar and vector mesons, such as the $D$ meson and its
vector partner, the $D^*$ meson. Thus, we calculate the self-energy and, hence, the spectral function of the $D$ and $D^*$ mesons in nuclear matter  from a  simultaneous self-consistent calculation in coupled channels that incorporates HQS. The $SU(3)$ WT meson-baryon lagrangian was extended to the $SU(8)$ spin-flavor symmetry group  including pseudoscalars and vector mesons together with $J=1/2^+$ and $J=3/2^+$ baryons \cite{magas09,gamermann10}. The $SU(8)$ symmetry is strongly broken in nature and this is incorporated  by adopting the physical hadron masses and different weak non-charmed and charmed pseudoscalar and vector meson decay constants. We also improve on the regularization scheme in nuclear matter going beyond the usual cutoff scheme \cite{tolos09}.

All resonances in the $SU(4)$ model are reproduced in the $SU(8)$ scheme and new resonant states are generated \cite{magas09} due to the enlarged Fock space. However, the nature of some of those resonances is different regarding the model. While the $\Lambda_c(2595)$ emerges as a $DN$ quasibound state in the $SU(4)$ model, it becomes predominantly a $D^*N$ quasibound state in the $SU(8)$ scheme.

The modifications of these resonances in the nuclear medium strongly depend on the coupling to $D$, $D^*$ and $N$ and are reflected in the spectral functions. On the r.h.s of  Fig.~\ref{fig1} we display the $D$ and $D^*$ spectral functions, which show then a rich spectrum of resonance-hole states. The $D$ meson quasiparticle peak mixes strongly with new resonant states $\Sigma_c(2823)N^{-1}$
and $\Sigma_c(2868)N^{-1}$ states while the $\Lambda_c(2595)N^{-1}$ is
clearly visible in the low-energy tail. The $D^*$ spectral function
incorporates the $J=3/2$ resonances, and the quasiparticle peak fully mixes with the new $\Sigma_c(2902)N^{-1}$ and $\Lambda_c(2941)N^{-1}$ states.  As density increases, these $Y_cN^{-1}$ modes tend to smear out and the
spectral functions broaden with increasing phase space, as seen before in the $SU(4)$ model \cite{mizutani06}.

\begin{figure}
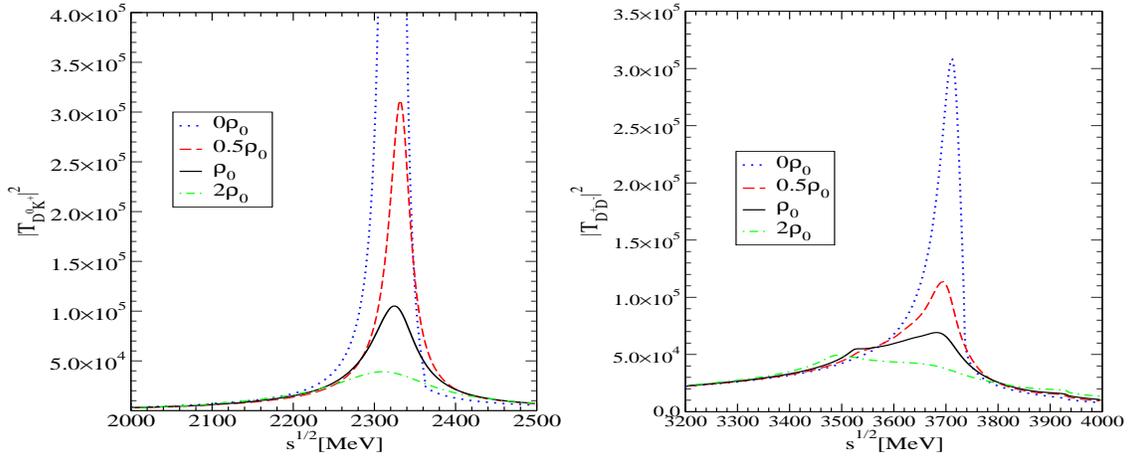

\includegraphics[width=0.5\textwidth,height=6cm]{ds02317}
\hfill
\includegraphics[width=0.5\textwidth,height=6cm]{x37}
\caption{$D_{s0}(2317)$ (left) and  $X(3700)$ (right) resonances in nuclear matter at zero temperature. \label{fig2}}
\end{figure}


\section{Hidden charm and charm scalar resonances \hspace{4cm} in nuclear matter}

The analysis of the properties of scalar resonances in nuclear matter is of great importance in order to understand their nature, whether they are $q \bar q$, molecules, mixtures of $q \bar q$ with meson-meson components, or dynamically generated resonances from the interaction of two pseudoscalars or two vectors. 

In the following we study the charmed resonance $D_{s0}(2317)$ \cite{Kolomeitsev:2003ac,guo06,Gamermann:2006nm} together with a hidden charm scalar meson, $X(3700)$, predicted in Ref.~\cite{Gamermann:2006nm}, which might have been observed by the Belle collaboration \cite{Abe:2007sy} via the reanalysis of Ref.~\cite{Gamermann:2007mu}. Those resonances are generated dynamically solving the coupled-channel Bethe-Salpeter equation for two pseudoscalars \cite{Molina:2008nh}. The kernel is derived from a $SU(4)$ extension of the $SU(3)$ chiral Lagrangian used to generate scalar resonances in the light sector. The $SU(4)$ symmetry is, however, strongly 
 broken, mostly due to the explicit consideration of the masses of the vector 
 mesons exchanged between pseudoscalars \cite{Gamermann:2006nm}. 

The transition amplitude around each resonance for the different coupled channels gives us information about the coupling of this state to a particular channel. The $D_{s0}(2317)$ mainly couples to the $DK$ system, while the hidden charm state $X(3700)$ couples most strongly to $D\bar{D}$. Thus, while the $K$ and $D$ self-energies are small compared to their mass, any change in the $D$ meson properties in nuclear matter will have an important effect on these  resonances. Those modifications are given by the $D$ meson self-energy in the $SU(4)$ model without the phenomenological isoscalar-scalar term, but supplemented by the $p$-wave self-energy through the corresponding $Y_cN^{-1}$ excitations \cite{Molina:2008nh}.

 In Fig.~\ref{fig2} the $D_{s0}(2317)$ and $X(3700)$ resonances are displayed via the squared transition amplitude for the corresponding dominant channel at different nuclear densities. The $D_{s0}(2317)$ and $X(3700)$ resonances, which have a zero and small width,
develop widths of the order of 100 and 200 MeV at normal nuclear matter density,  respectively. This is due to the opening of new many-body decay channels as the $D$ meson gets absorbed in the nuclear medium via $DN$ and $DNN$ inelastic reactions. We do not extract any clear conclusion for the mass shift. We suggest to look at transparency ratios to investigate those in-medium widths. This magnitude, which gives the survival probability in production reactions in  nuclei, is very sensitive to the in-medium width  of the resonance \cite{Hernandez:1992rv,Kaskulov:2006zc}.

\begin{figure}
\centerline{\includegraphics[width=0.43\textwidth,angle=-90]{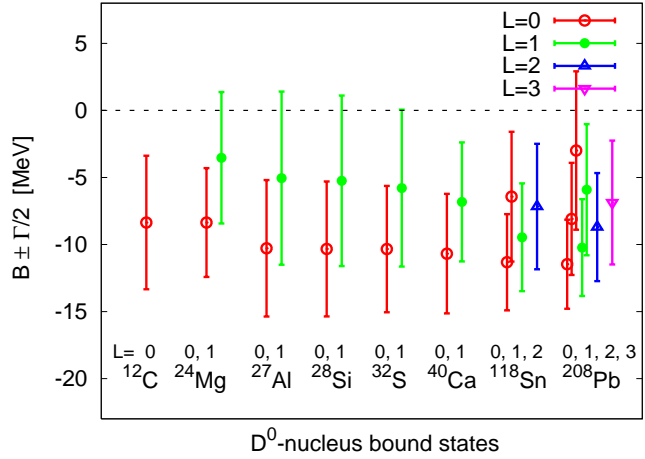}}
\caption{$D^0$ nucleus bound states with defined angular momentum $L$. \label{fig3}}
\end{figure}


\section{D-mesic nuclei}

The possible formation of $D$-meson bound states in $^{208}$Pb was predicted \cite{tsushima99} relying upon an attractive  $D$ meson potential in the nuclear medium  based on a quark-meson coupling (QMC) model \cite{sibirtsev99}. The experimental observation of those bound states, though, might be problematic since, even if there are bound states, their widths could be very large compared to the
 separation of the levels. This is indeed the case for the potential derived from a $SU(4)$ $t$-vector meson exchange model \cite{TOL07}.
However, the model that incorporates heavy-quark symmetry in the charm sector \cite{tolos09} generates widths of the $D$ meson in nuclear matter sufficiently small with respect to the mass shift to form bound states for $D$ mesons in nuclei.

In order to compute de $D$-nucleus bound states, we solve the Schr\"odinger equation. We concentrate on $D^0$-nucleus bound states since the Coulomb interaction prevents the formation of observable bound states for $D^+$ mesons. The potential that enters in the equation is an energy-dependent one that results from the zero-momentum $D$-meson self-energy within the SU(8) model \cite{tolos09}. In Fig.~\ref{fig3} we show  $D^0$ meson bound states in different nuclei. We observe that the $D^0$-nucleus states are weakly bound, in contrast to previous results using the QMC model, with significant widths \cite{carmen10}, in particular, for $^{208}$Pb \cite{tsushima99}.

\begin{figure}
\centerline{\includegraphics[width=0.43\textwidth]{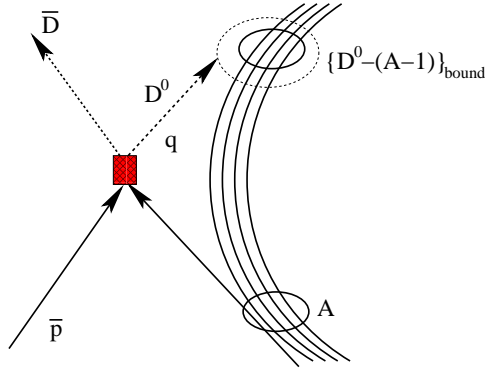}}
\caption{Possible mechanism for production of
  $D^0$ mesic nuclei with an antiproton beam\label{fig4}}
\end{figure}


The information on bound states is very valuable to gain some knowledge on the $D$ nucleus interaction, which is of interest for PANDA at FAIR. The experimental detection of $D^0$ meson bound states is, though, a difficult task. Reactions with antiprotons on nuclei (see Fig.~\ref{fig4}) might have a very low production rate. Similar reactions but with proton beams, although difficult, seem more likely to trap a $D^0$ in nuclei \cite{carmen10}.

\section{Conclusions and Outlook}

The properties of open-charm mesons ($D$ and $D^*$) in dense matter have been studied within a self-consistent coupled-channel approach taking, as bare interaction, different effective lagrangians. The in-medium solution  accounts for Pauli blocking effects and meson self-energies. We have analyzed the evolution with density and temperature of the open-charm meson spectral functions and discussed the implications of the properties of charmed mesons on the  $D_{s0}(2317)$ and the predicted $X(3700)$ in nuclear matter, and suggested to look at transparency ratios to investigate the changes in width of those resonances in nuclear matter. We have finally analyzed the possible formation of $D$-mesic nuclei. Only  weakly bound $D^0$-nucleus states seem to be feasible within the SU(8) scheme that incorporates heavy-quark symmetry. However, its experimental detection is most likely a challenging task.

\begin{theacknowledgments}
L.T. acknowledges support from the RFF program of the University of Groningen. This work is partly supported by the EU contract No. MRTN-CT-2006-035482 (FLAVIAnet), by the contracts FIS2008-01661 and FIS2008-01143 from MICINN (Spain), by the Spanish Consolider-Ingenio 2010 Programme CPAN (CSD2007-00042), by the Generalitat de Catalunya contract 2009SGR-1289 and by Junta de Andaluc\'{\i}a under contract FQM225. We acknowledge the support of the European Community-Research Infrastructure Integrating Activity ``Study of Strongly Interacting Matter'' (HadronPhysics2, Grant Agreement n. 227431) under the 7th Framework Programme of EU.
\end{theacknowledgments}


\bibliographystyle{aipproc}

\end{document}